\documentclass[useAMS,usenatbib]{mn2e}

\usepackage{graphicx}

\renewcommand{\ln}{\mathrm{ln}}

\title[On the central core in MHD winds and jets]
{On the central core in MHD winds and jets}
\author[V. S. Beskin and E. E. Nokhrina]{V. S. Beskin$^{1}$\thanks{E-mail:
beskin@lpi.ru} and E. E. Nokhrina$^{1,2}$
\\
$^{1}$P.N.Lebedev Physical Institute, Leninsky prosp., 53, Moscow, 119991, Russia\\
$^{2}$Moscow Institute of Physics and Technology, Dolgoprudny,
Moscow region, 141700, Russia}

\begin{document}

\date{Accepted, Received}

\pagerange{\pageref{firstpage}--\pageref{lastpage}} \pubyear{2008}

\maketitle

\label{firstpage}

\begin{abstract}
We demonstrate that the 1D cylindrical version of the
Grad-Shafranov equation is more rich than classical self-similar
ones, and more suitable for the astrophysical jets we observe. In
particular, it allows us to describe the central (and, hence, the
most energetic) part of the flow. Both relativistic and
non-relativistic versions are discussed. It is shown that taking
into account the finite pressure of the external media one can
determine the magnetic flux within the central core. We found as
well that for non-relativistic flows which are magnetically
dominated near the origin the solution can be constructed only in
the presence of the oblique shock near the base of a jet where the
additional heating is to take place.
\end{abstract}

\section{Introduction}\label{aba:sec1}

An activity of many compact objects -- Active Galactic Nuclei (AGN), Young Stellar
Objects (YSO), microquasars -- is associated with the highly collimated jets. These jets
are thought to be a natural outlet of an excess angular momentum of a central
object and accreting matter \citep{Heyvaerts-96}. The latest observations indicating
the jet rotation in AGN \citep{Y07} and YSO \citep{Bacciotti-07} support this idea.
The most attractive model for such outflows is the
MHD one \citep{Heyvaerts-96, bp, pp}.

Of course, the main question within this model is the collimation
itself \citep{bp, pp, st, Shu-94, op}. We assume here that the
collimation is due to a finite external gas and/or magnetic
pressure \citep{Appl-Camenzind-93, Lery-99, Beskin-Malyshkin-00}.
Indeed, proposing that it is the external magnetic field $B_{\rm
ext} \sim 10^{-6}$ G that plays the main role in the collimation,
we obtain $r_{\rm jet} \sim R_{\rm in}\left(B_{\rm in}/B_{\rm
ext}\right)^{1/2}$. Here $r$ is the distance from the rotational
axis, and the subscripts 'in' correspond to the values taken in
the vicinity of the central object. The similar evaluation can be
obtained for the external pressure $p_{\rm ext} \sim B^2_{\rm
ext}/8\pi$. As for YSO $B_{\rm in} \sim 10^3$ G and $R_{\rm in}
\sim R_{\odot}$, we obtain $r_{\rm jet} \sim 10^{15}$ cm, in
agreement with the observational data. Accordingly, for AGN
(\mbox{$B_{\rm in} \sim 10^4$ G}, $R_{\rm in} \sim 10^{13}$ cm) we have
$r_{\rm jet} \sim 1$ pc. It means that the external media may
indeed play an important role in the collimation process.

The internal structure of cylindrical jets was considered both for
non-relativistic \citep{cl, hn2} and relativistic
\citep{clb, Appl-Camenzind-93, eichler, bg2, ld, b3, bn} flows.
In particular, it was shown that for the constant angular velocity
of plasma $\Omega_{\rm F}$ it is impossible to obtain a reasonable
solution with total zero electric current
\citep{Appl-Camenzind-93}, but it can be constructed if the
angular velocity vanishes at the jet boundary and if the external
pressure is not equal to zero \citep{b3, Beskin-Malyshkin-00}.

Another result, obtained for both relativistic and
non-relativistic cylindrical flows \citep{clb, eichler, bg1, hn2},
is that the poloidal magnetic field $B_{\rm p}$ has a jet-like
form
\begin{equation}
B_{\rm p} = \frac{B_0}{1 + r^{2}/r_{\rm core}^2},
\label{r-2}
\end{equation}
where $r_{\rm core} = v_{\rm in}\gamma_{\rm in}/\Omega$. But this
relation corresponds to a very slow (logarithmic) growth of the
magnetic flux function: $\Psi(r) \propto \ln \, r$. It means that
if the jet core contains only a small part of the total magnetic
flux $\Psi_0$, the  jet boundary is located exponentially far from
the axis, with the magnetic field being too weak to be in the
equilibrium with the external pressure. In what follows we try to
resolve this contradiction.

Thus, we consider the following model: the flow crosses all
the critical surfaces while the effects of the external media are
negligible. It allows us to use standard values of
integrals of motion. As the supersonic wind expands, its pressure
becomes comparable with the external gas and/or magnetic pressure.
The interaction of a flow with external media results in well collimated
jet which can be described by 1D cylindrical equations.

\section{Basic equations}

\subsection{Relativistic flow}

For a cylindrical flow one can write down electric ${\bf E}$ and
magnetic ${\bf B}$ fields as well as the four-velocity of a
plasma ${\bf u}$ in the standard form
\begin{equation}
 {\bf B} = \frac{{\bf \nabla}\Psi \times {\bf e}_{\varphi}}{2\pi r}
  -\frac{2I}{r c}{\bf e}_{\varphi}, \qquad
{\bf E} =  -\frac{\Omega_{\rm F}}{2\pi c}{\bf \nabla}\Psi,
\label{defE}
\end{equation}
\begin{equation}
 {\bf u}  = \frac{\eta}{n}{\bf B}
+ \gamma \left(\frac{\Omega_{\rm F}r}{c}\right) {\bf e}_{\varphi}.
  \label{4a''}
\end{equation}
Here $n$ is the concentration in the comoving reference frame, and
$\gamma^2 = {\bf u}^2 + 1$ is the Lorentz-factor. In other words,
it is convenient to represent all the values in terms of a
magnetic flux $\Psi$ and a total electric current $I$, the angular
velocity of plasma $\Omega_{\rm F}$ and the particle to magnetic
flux ratio $\eta$ being constant on the magnetic surfaces:
$\Omega_{\rm F} = \Omega_{\rm F}(\Psi)$, $\eta = \eta(\Psi)$.
Accordingly, the trans-field Grad-Shafranov (GS) equation can be rewritten as
\citep{b4}
\begin{eqnarray}
\frac{1}{r}\frac{{\rm d}}{{\rm d} r }
\left(\frac{A}{r}\frac{{\rm d}\Psi}{{\rm d} r }\right)
+\frac{\Omega_{\rm F}}{c^2}\left(\frac{{\rm d}\Psi}{{\rm d} r }\right)^2
\frac{{\rm d}\Omega_{\rm F}}{{\rm d}\Psi}
+\frac{32\pi^4}{r ^2{\cal M}^2 c^4}
\frac{{\rm d}}{{\rm d}\Psi}\left(\frac{G}{A}\right) \nonumber \\
-\frac{64\pi^4\mu^2}{{\cal M}^2}\eta\frac{{\rm d}\eta}{{\rm
d}\Psi} - 16\pi^{3}n T \frac{{\rm d}s}{{\rm d}\Psi}=0. \label{ap1}
\end{eqnarray}
Here the entropy $s = s(\Psi)$ is the fifth integral of motion,
\begin{equation}
G = r ^2(E-\Omega_{\rm F} L)^2 +{\cal M}^{2}L^2c^2-{\cal M}^2 r ^{2} E^2,
\end{equation}
$A = 1-\Omega_{\rm F}^2 r ^2/c^2-{\cal M}^2$
is the Alfv\'enic factor,
${\cal M}^2 = 4\pi\mu\eta^2/n$
is the poloidal Alfv\'enic Mach number,
$\mu=m_{\rm p}c^2+m_{\rm p}w$
is the relativistic enthalpy, and the derivative ${\rm d}/{\rm
d}\Psi$ acts on the integrals of motions only. Finally, the
relativistic Bernoulli equation $u_{\rm p}^2 = \gamma^2 -
u_{\varphi}^2 - 1$ has a form
\begin{eqnarray}
\frac{{\cal M}^4}{64\pi^4 r^2}
\left(\frac{{\rm d}\Psi}{{\rm d}r}\right)^2 =
\frac{K}{r^2A^2c^4 } - \mu^2\eta^2,
\label{ap4}
\end{eqnarray}
where
\begin{equation}
K= r^2(e')^2(A-{\cal M}^2)
+{\cal M}^4 r^{2}E^2-{\cal M}^{4}L^2c^2,
\label{ap5}
\end{equation}
and $e' = E-\Omega_{\rm F}L$.
Both equations contain relativistic integrals of motion
\begin{eqnarray}
E(\Psi) = \gamma\mu\eta c^2 + \frac{\Omega_{\rm F}I}{2\pi},  \quad
L(\Psi) = r u_{\varphi}\mu\eta c + \frac{I}{2\pi},
\label{ap7}
\end{eqnarray}
which, as all other invariants, are to be determined from
boundary and critical conditions. E.g., for the inner part of a
flow $\Psi \ll \Psi_0$ with a zero temperature one can choose
$\Omega_{\rm F}(\Psi)= \Omega_0$, $\eta(\Psi)=\eta_0$, and
\begin{eqnarray}
E(\Psi)= \mu\eta_0\gamma_{\rm in}c^2 + \frac{\Omega_0^2}{4\pi^2}\Psi, \quad
L(\Psi)= \frac{\Omega_0}{4\pi^2}\Psi.
\label{ProbStat-L1}
\end{eqnarray}

Multiplying now equation (\ref{ap1}) by $2A{\rm d}\Psi/{\rm d}
r $ and using equation (\ref{ap4}), one can find \citep{b3}
\begin{eqnarray}
\left[\frac{(e')^2}{\mu^2\eta^2}-1+\frac{\Omega_{\rm F}^2 r^2}{c^2}
-A\frac{c_s^2}{c^2}\right]
\frac{{\rm d}{\cal M}^2}{{\rm d} r } =  \nonumber \\
\frac{{\cal M}^6L^2}{A r ^3 \mu^2\eta^2c^2}
+\frac{\Omega_{\rm F}^2 r {\cal M}^2}{c^{2}}\left[2 - \frac{(e')^2}{A\mu^2\eta^2c^4}\right]
\nonumber \\
+{\cal M}^2 \frac{e'}{\mu^2\eta^2c^4}\frac{{\rm d}\Psi}{{\rm d}r}\frac{{\rm d}e'}{{\rm d}\Psi}
+{\cal M}^2\frac{ r ^2}{c^2}\Omega_{\rm F}\frac{{\rm d}\Psi}{{\rm d}r}
\frac{{\rm d}\Omega_{\rm F}}{{\rm d}\Psi}
\label{ap10} \\
-{\cal M}^2 \left(1-\frac{\Omega_{\rm F}^2 r ^2}{c^2} + 2A\frac{c_s^2}{c^2}\right)
\frac{{\rm d}\Psi}{{\rm d} r }\frac{1}{\eta}\frac{{\rm d}\eta}{{\rm d}\Psi} \nonumber \\
-\left[\frac{A}{n}\left(\frac{\partial P}{\partial s}\right)_n
+\left(1-\frac{\Omega_{\rm F}^2 r ^2}{c^2}\right)T\right]
\frac{{\cal M}^2}{\mu}\frac{{\rm d}\Psi}{{\rm d}r}
\frac{{\rm d}s}{{\rm d}\Psi},
\nonumber
\end{eqnarray}
where $T$ is the temperature and $c_s \ll c$ is the sound velocity. Together with the
Bernoulli equation (\ref{ap4}) it forms the system of two
ordinary differential equations for the Mach number ${\cal M}^2(r)$
and the magnetic flux $\Psi(r)$ describing cylindrical relativistic
jet. Clear boundary conditions are
\begin{eqnarray}
\Psi(0) = 0, \quad
P(r_{\rm jet}) = P_{\rm ext},
\end{eqnarray}
where $P = B^2/8\pi + p$ is the total pressure. Determining the
functions ${\cal M}^2(r)$ and $\Psi(r)$, one can find the jet
radius $r_{\rm jet}$ as well as the profile of the current $I$,
the particle energy, and the toroidal component of the
four-velocity using standard expressions
\begin{eqnarray}
\frac{I}{2\pi} & = & \frac{L-\Omega_{\rm F}r^{2}E/c^2}
{1-\Omega_{\rm F}^{2}r^{2}/c^2-{\cal M}^{2}},
\label{p33} \\
\nonumber \\
\gamma & = & \frac{1}{\mu\eta c^2} \, \frac{(E-\Omega_{\rm F}L)-{\cal M}^{2}E}
{1-\Omega_{\rm F}^{2}r^{2}/c^2-{\cal M}^{2}},
\label{p34} \\
\nonumber \\
u_{\varphi} & = & \frac{1}{\mu\eta r c} \, \frac{(E-\Omega_{\rm F}L)
 \Omega_{\rm F}r^{2}/c^2-L{\cal M}^{2}}{1-\Omega_{\rm F}^{2}r^{2}/c^2-{\cal M}^{2}}.
\label{p35}
\end{eqnarray}

\subsection{Non-relativistic flow}

In the non-relativistic limit the electric and magnetic fields are
determined by general expressions (\ref{defE}). On the other
hand, equation (\ref{4a''}) can be rewritten as
\begin{eqnarray}
 {\bf v} & = & \frac{\eta_{\rm n}}{\rho_{\rm m}}{\bf B}+\Omega_{\rm F}
  r{\bf e}_{\varphi},
  \label{4a}
\end{eqnarray}
where $\rho_{\rm m} = m_{\rm p}n$ is the mass density, and
$\eta_{\rm n}(\Psi)$ is the non-relativistic particle to magnetic
flux ratio. Accordingly, non-relativistic fluxes of energy
$E_{\rm n}$ and $z$ component of the angular momentum $L_{\rm n}$
are
\begin{eqnarray}
  E_{\rm n}(\Psi) & = & \frac{\Omega_{\rm F} I}{2\pi c\eta_{\rm n}}
  +\frac{v^2}{2} + w,
  \label{5a}  \\
   L_{\rm n}(\Psi) & = & \frac{I}{2\pi c\eta_{\rm n}}+v_{\varphi}r.
    \label{6a}
\end{eqnarray}
Further, algebraic relations (\ref{p33}), (\ref{p35}) can be rewritten
as
\begin{eqnarray}
 \frac{I}{2\pi} & = & c\eta_{\rm n}
\frac{L_{\rm n}-\Omega_{\rm F}r^2}{1-{\cal M}^2},
\label{Inrel} \\
   v_{\varphi} & = & \frac{1}{r}\frac{\Omega_{\rm F}r^2
    -L_{\rm n}{\cal M}^2}{1-{\cal M}^2},
    \label{9a}
     \end{eqnarray}
where now
\begin{equation}
 {\cal M}^{2}=\frac{4\pi\eta_{\rm n}^{2}}{\rho_{\rm m}}.
  \label{10a}
   \end{equation}
As a result, the non-relativistic Bernoulli equation
\begin{eqnarray}
  \frac{{\cal M}^4}{64\pi^4\eta_{\rm n}^2}\left(\frac{{\rm d}\Psi}{{\rm d}r}\right)^2
    = 2r^{2} (E_{\rm n} - w) \nonumber \\
  -\frac{(\Omega_{\rm F}r^2- L_{\rm n}{\cal M}^2)^2}
{(1-{\cal M}^2)^2}
      -2r^2\Omega_{\rm F}\frac{L_{\rm n}-\Omega_{\rm F}r^2}
{1-{\cal M}^2},
       \label{11a}
        \end{eqnarray}
together with the non-relativistic limit of equation (\ref{ap10})
\begin{eqnarray}
\displaystyle\left[\vphantom{\frac{1}{2}}2e_{\rm n}-
2w+\Omega_{\rm F}^2r^2-(1-{\cal M}^2)c_{\rm s}^2\right] \frac{{\rm d}{\cal M}^2}{{\rm d}r} = \nonumber \\
\displaystyle \frac{{\cal M}^6}{1-{\cal M}^2}\frac{L_{\rm n}^2}{r^3}
- \frac{\Omega_{\rm F}^2r}{1-{\cal M}^2}{\cal M}^2(2{\cal M}^2-1)  \nonumber \\
\displaystyle  + {\cal M}^2\frac{{\rm d}\Psi}{{\rm d}r}\frac{{\rm d}e_{\rm n}}{{\rm d}\Psi}
+{\cal M}^2r^2\Omega_{\rm F}\frac{{\rm d}\Psi}{{\rm d}r} \frac{{\rm d}\Omega_{\rm F}}{{\rm d}\Psi} \nonumber \\
\displaystyle  +2 \left[e_{\rm n}-w+\frac{\Omega_{\rm F}^2r^2}{2}-(1-{\cal M}^2)c_{\rm s}^2\right]
\frac{{\cal M}^2}{\eta_{\rm n}}\frac{{\rm d}\Psi}{{\rm d}r}\,\frac{{\rm d}\eta_{\rm n}}{{\rm d}\Psi} \nonumber \\
\displaystyle -{\cal M}^2\left[(1-{\cal M}^2)\frac{1}{\rho_{\rm m}}
\left(\frac{\partial P}{\partial s}\right)_{\rho_{\rm m}}
+\frac{T}{m_{\rm p}}\right]\frac{{\rm d}\Psi}{{\rm d}r} \frac{{\rm {\rm d}}s}{{\rm d}\Psi},
\label{GS-Me}
\end{eqnarray}
where $e_{\rm n} = E_{\rm n}-\Omega_{\rm F}L_{\rm n}$, determine
the structure of a non-relativistic cylindrical flow.

\section{Advantages}

Certainly, the approach under consideration is one-dimensional.
For this reason, it has some properties similar to the other
self-similar ones. In particular, one can easily check that the
singularity on the fast magnetosonic surface is absent. On the
other hand, singularity appears on the cusp surface where the
factors in front of ${\rm d}{\cal M}^2/{\rm d} r$ in (\ref{ap10})
and (\ref{GS-Me}) vanish. Nevertheless, in our opinion, this
one-dimensional approach has some clear advantages in comparison
with the standard self-similar ones \citep{bp, pp, st, Shu-94}.

First of all, it allows us to use any form of the five integrals of motion.
Indeed, the self-similarity of a flow demands definite dependence of invariants
which may not correspond to the real boundary conditions. E.g.,
for relativistic self-similar flow the angular velocity
$\Omega_{\rm F}$ is to have the form $\Omega_{\rm F}\propto
r^{-1}$ \citep{lcb}. It does not correspond neither to the
homogeneous stellar rotation, nor to the Keplerian disk rotation.
Moreover, this dependence has the singularity at the rotational
axis. Thus, the standard self-similar approach cannot describe
the central (and, hence, the most energetic) part of the flow.

Further, classical self-similar approach cannot describe the
region of the electric current closure. Finally, for the
relativistic magnetically dominated flow it is more convenient to
use first-order equation (\ref{ap10}) instead of second order GS
equation for which it is necessary to be careful in taking into
account small but important terms $\sim \gamma^{-2}$. Indeed, the
force balance equation (\ref{ap10}) does not contain
the leading terms $\rho_{\rm e}{\bf E}$ and ${\bf j} \times {\bf B}/c$ as they
are analytically removed using Bernoulli equation.
As a result, as
\begin{equation}
\frac{|\rho_{\rm e}{\bf E} + {\bf j} \times {\bf B}/c|}
{|{\bf j} \times {\bf B}/c|} \sim \frac{1}{\gamma^2},
\label{forcebls}
\end{equation}
all the terms in equation (\ref{ap10}) are of the same order.

In particular, in the limit $r \gg r_{\rm core}$, ${\cal M}^2 \gg 1$ equation
(\ref{ap10}) can be rewritten in the simple form \citep{Beskin-Malyshkin-00}
\begin{equation}
\frac{{\rm d}}{{\rm d}r}
\left(\frac{\mu\eta\Omega_{\rm F}r^2}{{\cal M}^2}\right)
- \frac{{\cal M}^2}
{\mu\eta \Omega_{\rm F} r^3(\Omega_{\rm F}^2r^2/c^2 + {\cal M}^2)}L^2 = 0.
\label{qq}
\end{equation}
Without the last term $\propto L^2(\Psi)$ equation (\ref{qq}) results in the
conservation of the value $H$
\begin{equation}
H = \frac{\Omega_{\rm F}\eta r^2}{{\cal M}^2} =  {\rm const},
\label{qq1}
\end{equation}
was found by \citet{Heyvaerts-Norman-89} for the conical
magnetic field. It is the conservation of $H$ that results in the
jet-like solution (\ref{r-2}). Indeed, as $\eta(\Psi) \approx $
const and $\Omega_{\rm F}(\Psi) \approx$ const in the very center
of a jet, we obtain ${\cal M}^2 \propto r^2$. Using now the
definitions ${\cal M}^2 = 4\pi \eta^2\mu /n$ and $n u_{\rm p} =
\eta B_{\rm p}$ (and the condition $u_{\rm p} \approx$ const in
the very center of a flow), we return to (\ref{r-2}). But, as we
will see, the term containing $L^2$ (which appears to be missed
previously) can be important \citep{b3}. It is this term that can
change the jet-like structure in a relativistic case.

\section{Internal structure of cylindrical jets}

\subsection{Relativistic cold flow}

\subsubsection{General properties}

The solution of equations (\ref{ap4}) and (\ref{ap10})
for relativistic cold flow $c_{\rm s} = 0$, $s = 0$ depends
essentially on the Mach number on the rotational axis ${\cal
M}_0^2 = {\cal M}^2(0)$ \citep{Beskin-Malyshkin-00}. For ${\cal
M}_0^2 \gg {\cal M}_{\rm cr}^2$, where
\begin{equation}
{\cal M}_{\rm cr}^2 = \gamma_{\rm in}^2
\label{AAA}
\end{equation}
we have
\begin{equation}
{\cal M}^{2} =
{\cal M}_{0}^{2}\left(1 + \frac{r^2}{\gamma_{\rm in}^2 R_{\rm L}^2}\right),
\label{sol0}
\end{equation}
the poloidal magnetic field corresponding to jet-like solution (\ref{r-2}).
Here $R_{\rm L} = c/\Omega_{\rm F}(0)$.
On the other hand, for ${\cal M}_0^2 \ll {\cal M}_{\rm cr}^2$ we obtain
\begin{eqnarray}
\Psi & = & \frac{\gamma_{\rm in} \Psi_0}{2{\cal M}_{0}^{2}\sigma}
\left(\frac{r}{R_{\rm L}}\right)^2,
\label{sol1'} \\
{\cal M}^{2} & = & {\cal M}_{0}^{2}
\left(1 + \frac{r}{\gamma_{\rm in} R_{\rm L}}\right).
\label{sol1}
\end{eqnarray}
Here
\begin{equation}
\sigma = \frac{\Omega_{0}^2\Psi_{0}}{8\pi^2 c^2 \mu\eta_{0}}
\label{sigma}
\end{equation}
is the Michel
magnetization parameter \citep{michel69} ($\gamma \approx \sigma$ for
particle dominated flow $W_{\rm part} \approx W_{\rm em}$).
It means that $B_{\rm p} \approx$ const, i.e., the solution has no
jet-like form.

As was already stressed, the solution (\ref{sol0}) cannot be
realized in the presence of the external media. Hence, one
can conclude that for any finite external pressure $P_{\rm ext}$
magnetic field in the center of cylindrical jet
$B_0 = 4 \pi \eta \mu \gamma_{\rm in}/{\cal M}_{0}^2$ cannot be much
smaller than
$B_{\rm min} = 4 \pi \eta \mu \gamma_{\rm in}/{\cal M}_{\rm cr}^2$.
It gives
\begin{equation}
B_{\rm min} = \frac{1}{\sigma\gamma_{\rm in}}B(R_{\rm L}),
\label{38}
\end{equation}
where $B(R_{\rm L}) = \Psi_0/\pi R_{\rm L}^2$.

\subsubsection{Central core}

Thus, for the external magnetic field $B_{\rm ext} > B_{\rm min}$
the internal structure of a relativistic jet is to be described
by relations (\ref{sol1'})--(\ref{sol1}) when $B_{\rm p} \approx B_{\rm ext}$.
On the other hand, for
$B_{\rm ext} <B_{\rm min}$ in the center of a flow (i.e., for
$r < \gamma_{\rm in} R_{\rm L}$) the core with
$B_{\rm p} \approx B_{\rm min}$ is
formed. As was found \citep{bn}, for $\sigma^{-2}B(R_{\rm L}) <
B_{\rm ext} < B_{\rm min}$ (and for $r \gg \gamma_{\rm in} R_{\rm
L}$) the solution can be presented as
\begin{equation}
\Psi \propto r^{a} , \quad {\cal M}^{2} \propto r^{b},
\label{a+b}
\end{equation}
the sum being $a + b = 3$. E.g., for $B_{\rm ext} = B_{\rm min}$
we have $a = 2$, $b = 1$ (cf. (\ref{sol1'})--(\ref{sol1})), and for
$B_{\rm ext} = \sigma^{-2}B(R_{\rm L})$ we have $a = 1$, $b = 2$.

The results presented above were reproduced recently both
analytically and numerically. In \citep{bn} it was shown that 1D
approximation becomes true for the paraboloidal outflow at large
distances from the equatorial plane $z \gg \sigma^{2/3}R_{\rm L}$
where the flow becomes actually cylindrical. Up to the distance
$z=\sigma\gamma_{\rm in}R_{\rm L}$ one can use the relations
(\ref{sol1'})--(\ref{sol1}), so the poloidal magnetic field does not depend on
$r$. The region $z > \sigma \gamma_{\rm in} R_{\rm L}$
corresponds to core-like solution (\ref{a+b}). Nevertheless, the
transverse dimension of a jet remains parabolic: $r_{\rm jet}
\propto z^{1/2}$. Numerically the scalings (\ref{a+b}) were
confirmed by \citet{kbvk}.

Remember that the existence of a cylindrical core with $r_{\rm
core} \sim \gamma_{\rm in }R_{\rm L}$ was predicted in many
papers \citep{Heyvaerts-Norman-89, bg2}, but the magnetic flux
\begin{equation}
\Psi_{\rm core} \approx \pi r_{\rm core}^2B_{\rm min}
\end{equation}
inside the core was unknown up to now. As we see, in the
relativistic case the central core contains only a small part of
the magnetic flux:
\begin{equation}
\frac{\Psi_{\rm core}}{\Psi_0} \approx \frac{\gamma_{\rm in}}{\sigma}.
\label{39}
\end{equation}
Nevertheless, as $b > 0$, such core-like flow can exist in
the presence of the external media.

\subsubsection{Bulk acceleration}

As on the fast magnetosonic surface ($r_{\rm F} \sim
\sigma^{1/3}R_{\rm L}$) in the region of the diverging magnetic
field lines the bulk plasma Lorentz-factor $\gamma(r_{\rm F}) =
\sigma^{1/3}$ \citep{michel69, bkr}, and, hence, here
\begin{equation}
\frac{W_{\rm part}}{W_{\rm em}} \sim \sigma^{-2/3} \ll 1,
\end{equation}
an additional particle acceleration is possible
as the transverse dimension of the diverging flow becomes larger
than $r_{\rm F}$. Using equation (\ref{p34}) and the relation
$a + b = 3$ one can find that in the whole region
$B_{\rm ext} > \sigma^{-2}B(R_{\rm L})$ ($z < \sigma^2 R_{\rm L}$
for the parabolic flow) the Lorentz-factor can be determined as
\begin{equation}
\gamma \approx r/R_{\rm L}.
\label{gx}
\end{equation}
Accordingly, one can write down
\begin{equation}
\frac{W_{\rm part}}{W_{\rm em}} \sim
\frac{1}{\sigma}\left[\frac{B(R_{\rm L})}{B_{\rm ext}}\right]^{1/2}.
\label{46b}
\end{equation}
It means that for  $B_{\rm ext} \sim \sigma^{-2}B(R_{\rm L})$ ($z
\sim \sigma^2 R_{\rm L}$ for the parabolic flow), where the
transverse jet dimension
\begin{equation}
r_{\rm jet} \sim \sigma R_{\rm L},
\end{equation}
almost the full energy transformation from the Poynting flux to
the particle energy flux can be realized. In particular, for the
particle moving along the parabolic magnetic field line one can
obtain
\begin{equation}
\gamma(z) \approx (z/R_{\rm L})^{1/2}.
\label{gz}
\end{equation}
This scaling was confirmed numerically as well \citep{McK, Narayan}.

It is necessary to stress that relation (\ref{gx}) takes place
only if one can neglect the curvature of the magnetic surfaces.
Indeed, for the magnetically dominated case in the limit $r \gg
r_{\rm F}$ the leading terms in two-dimensional GS equation
$\rho_{\rm e}{\bf E} + {\bf j} \times {\bf B}/c$ can be rewritten
in the simple form \citep{bn}
\begin{equation}
\frac{1}{2} \, {\bf n} \cdot \nabla (B_{\rm p}^2) -
\frac{{B_{\varphi}^2}}{R_{\rm c}}({\bf n} \cdot {\bf R}_{c}) +
\frac{B_{\varphi}^2 - {\bf E}^2}{r} \, ({\bf n} \cdot {\bf
e}_{r}) = 0. \label{balance}
\end{equation}
Here $R_{\rm c}$ is the (poloidal) curvature radius of magnetic
surfaces, ${\bf R}_{c}$ is the unit vector in the direction of
curvature radius growth, and ${\bf n} = \nabla \Psi/|\nabla \Psi|$.

Neglecting now the curvature term
and using standard relations $B_{\varphi} \approx B_{\rm p} r/R_{\rm L}$
and $B_{\varphi}^2 - {\bf E}^2 \approx B_{\varphi}^2/\gamma^2$
resulting from (\ref{defE}) and (\ref{ap4}), we return to (\ref{gx}). On
the other hand, if the curvature is important, then one can neglect the
first term in (\ref{balance}), and we obtain \citep{sol}
\begin{equation}
\gamma \approx \left(R_{\rm c}/r\right)^{1/2}.
\label{gx1}
\end{equation}
This scaling taking place for the split-monopole geometry outside
the fast magnetosonic surface corresponds to \citep{tomi, bkr}
\begin{equation}
\gamma \approx \sigma^{1/3}\ln^{1/3}(r/r_{\rm F}).
\label{gx2}
\end{equation}
Remember that for $r < r_{\rm F}$ we have a "linear" acceleration
(\ref{gx}).

As was demonstrated numerically \citep{Narayan}, it is the
parabolic flow that terminates these two asymptotic solutions. If the
magnetic surfaces have a form $z \propto r^{k}$ then for $k > 2$
(when the collimation is even stronger than for a parabolic flow)
one can use the relation $\gamma \approx r/R_{\rm L}$. On the
other hand, for $1 < k < 2$ at large distances the particles
acceleration is not so effective, so that
$\gamma \approx (R_{\rm c}/r)^{1/2}$. As $R_{\rm c} \approx (z')^3/z''$
for $z' \gg 1$, where $z' = {\rm d}z/{\rm d} r$, we obtain for the Lorentz-factor
of a particle moving along the magnetic field line $z=z(r)$
\begin{equation}
\gamma \propto z^{(k-1)/k},
\label{qq'}
\end{equation}
in full agreement with the cases $k = 2$ (\ref{gx1}) and $k = 2$
(\ref{gx2}) considered above.
Accordingly, the total energy transformation can be realized if the jet width is
\begin{equation}
r_{\rm eq} \sim \sigma^{1/(k-1)}R_{\rm L}.
\label{qqq}
\end{equation}
For $k = 3/2$ these scalings were confirmed numerically by \citet{bk08}.
For $k > 2$ the evaluation $r_{\rm eq} \sim \sigma R_{\rm L}$ is
to be used. Thus, effective particle acceleration can take place
only if $r_{\rm jet} \geq \sigma R_{\rm L}$, and if the curvature
of magnetic surfaces is not important.

\subsubsection{In the center of the self-similar solution}

The approach under consideration allows us matching the
self-similar solution to the rotational axis. Indeed, for the
self-similar invariants \citep{lcb,cl}
\begin{eqnarray}
\Omega_{\rm F}(\Psi) & = &  \Omega_0 (\Psi/\Psi_{\rm b})^{-\beta},
\label{int1''} \\
E(\Psi) & = & E_0 (\Psi/\Psi_{\rm b})^{1 - 2\beta},
\label{int2''} \\
L(\Psi) & = & L_0 (\Psi/\Psi_{\rm b})^{1 - \beta},
\label{int3''} \\
\eta(\Psi) & = & \eta_0 (\Psi/\Psi_{\rm b})^{1 - 2\beta},
\label{int4''}
\end{eqnarray}
one can seek the solution of two-dimensional GS equation for $\Psi > \Psi_{\rm b}$
in the form $\Psi(R,\theta) = R^{1/\beta} \Theta(\theta)$, where $R$
is the spherical radius. Hence, for $\theta \ll 1$ one can write
down
\begin{equation}
\Psi(R,\theta) = {\cal A} R^{1/\beta} \theta^a,
\label{autoan}
\end{equation}
where ${\cal A} =$ const, so the cylindrical radius of the boundary $\Psi = \Psi_{\rm b}$
can be written as
\begin{equation}
r_{\rm b}(z) = {\cal A}^{-1/a} \Psi_{\rm b}^{1/a} z^{1 - 1/ab}.
\end{equation}

On the other hand, as was already stressed, in the central part of a flow
$\Psi < \Psi_{\rm b}$ the self-similar approach cannot be used.
For simplicity we assume that $\Omega_{\rm F}  = \Omega_0 = $ const and
$\eta=\eta_0 =$ const
for $\Psi < \Psi_{\rm b}$. Then far from the equatorial plane where
$z \gg r$  ($\theta \ll 1$) one can integrate 1D cylindrical equations
(\ref{ap4}) and (\ref{ap10}) considering $z \approx R$ as a
parameter.  As a result, using solutions (\ref{sol1'}) and (\ref{sol1}),
we obtain for ${\cal M}^2_{\rm b}(z) = {\cal M}^2(r_{\rm b})$ two
different expressions for particle and magnetically dominated flows.

For magnetically dominated flow ($r_{b}\gg \gamma_{\rm in}R_{\rm L}$) we have
\begin{equation}
{\cal M}^2_{\rm b}(z) = \frac{4\pi^2 \eta_0 \mu}{R_{\rm L}{\cal
A}^{3/a} \Psi_{\rm b}^{1-3/a}} z^{3 -3/a\beta}.
\end{equation}
Besides, for $\Psi > \Psi_{\rm b}$ ($r > r_{\rm b}$)
one can seek the solution in a form
\begin{equation}
{\cal M}^2(r) = {\cal M_{\rm b}}^2(r/r_{\rm b})^{b}.
\label{Masy}
\end{equation}
As a result, equations (\ref{ap4}), (\ref{ap10}) give
\begin{equation}
a = 2, \quad b = 3 - 6\beta,
\label{asy1}
\end{equation}
the first relation demonstrating that poloidal magnetic
field is to be homogeneous: $B_{\rm p} \approx $ const.
Substituting now $r = R \theta$, we see that for $\Psi > \Psi_{\rm b}$
\begin{equation}
{\cal M}^2 = {\cal C} \theta^{b},
\end{equation}
where the coefficient
\begin{equation}
{\cal C} \propto {\cal M_{\rm b}}^2(R)R^{b}/r_{\rm b}^{b}(R)
\propto R^{3+(b-3)/2\beta},
\end{equation}
in agreement with
the self-similar property, does not depend on $R$.

\begin{figure}
\includegraphics{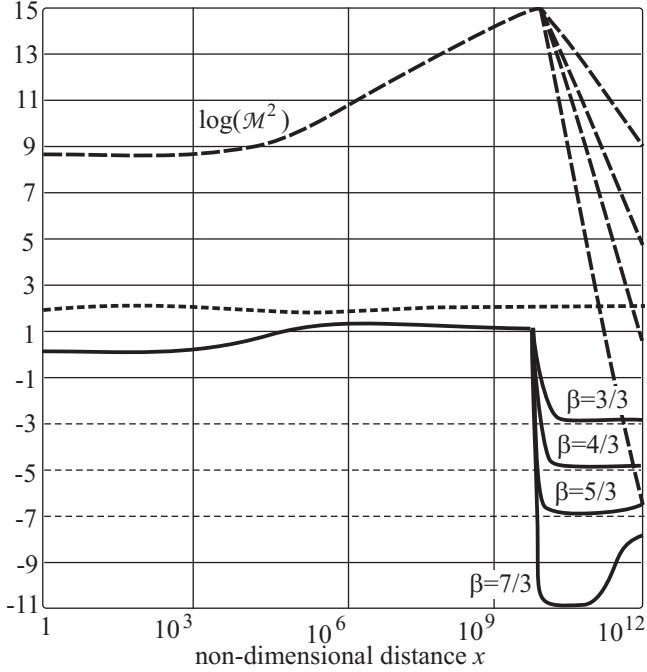}
\caption{Internal structure of magnetically dominated
relativistic jet. The short dashes line is an exponent $a$ of the
flux function $\Psi$ (\ref{autoan}). The long dashes line
represents the $\mathrm{log}{\cal M}^2$, and the solid lines
represent an exponent $b$ of ${\cal M}^2$ (\ref{Masy}) for
different values of the parameter $\beta$. The thin dashes lines
are the analytical exponents $b = 3 - 6\beta$. The
non-dimensional radius $x=\Omega_{\rm F}(0)r/c$.}
\end{figure}

For particle dominated flow ($r_{b}\ll \gamma_{\rm in}R_{\rm L}$) we find
\begin{equation}
{\cal M}^2_{\rm b}(z) = \frac{4\pi^2 \eta_0 \mu\gamma_{\rm in}}
{{\cal A}^{2/a} \Psi_{\rm b}^{1-2/a}} z^{2 - 2/a\beta}.
\end{equation}
Seeking again the solution for $\Psi>\Psi_{\rm b}$ in a form
(\ref{Masy}), we have
\begin{equation}
a = 2, \quad b = 2 - 4\beta.
\label{asy2}
\end{equation}
This solution ensures independence of a coefficient
${\cal C}$ on $R$ as well.
The results of numerical integration of the system
(\ref{ap4}), (\ref{ap10}) for the cold flow are presented
in Figs. 1, 2. As we see, there is very good agreement between numerical
results an analytical asymptotic behaviour (\ref{asy1}) and (\ref{asy2}).

\begin{figure}
\includegraphics{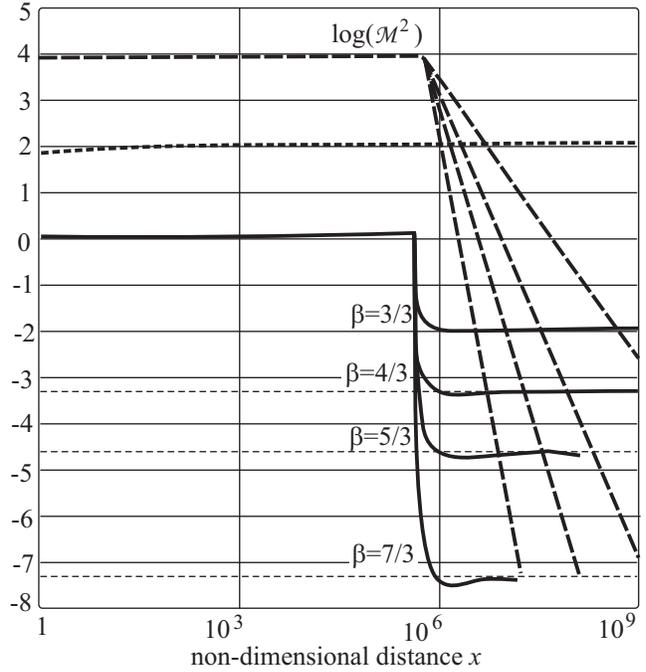}
\caption{The same for particle dominated relativistic jet.
The thin dashes lines
are the theoretical exponents $b = 2 - 4\beta$}
\end{figure}

\subsection{Non-relativistic flow}

\subsubsection{Central core}

For the non-relativistic cold flow in the central part of a jet one
can use general expressions (\ref{ProbStat-L1})
\begin{equation}
E_{\rm n}(\Psi)= \frac{v_{\rm in}^2}{2} + i_0\frac{\Omega_0^2}{4\pi^2 c\eta_0}\Psi, \quad
L_{\rm n}(\Psi)=i_0\frac{\Omega_0}{4\pi^2 c \eta_0}\Psi,
\label{ProbStat-L2}
\end{equation}
where $v_{\rm in}$ can be considered as a constant, and the
non-dimensional longitudinal current $i_0 = j/j_{\rm GJ}$ depends
now on the angular velocity $\Omega_{\rm F}$. Here $j_{\rm GJ} =
\Omega_0 B/2\pi$ is the Goldreich-Julian current density. For
$\Omega_{\rm F} \ll \Omega_{\rm cr}$, where
\begin{equation}
\Omega_{\rm cr} = \frac{v_{\rm in}}{R_{\rm in}}
\left( \frac{\rho_{\rm in}v_{\rm in}^2}{B_{\rm in}^2/8\pi}\right)^{1/2},
\end{equation}
corresponding to a particle dominated outflow near the star, the
2D problem can be solved analytically \citep{bg1, bo}, and we
obtain $i_0 = c/v_{\rm in}$. For a magnetically dominated flow
near the origin one can write down (see, e.g., \citep{Lery-99})
\begin{equation}
i_0 \approx \frac{c}{v_{\rm in}} \left(\frac{\Omega_{\rm F}}{\Omega_{\rm cr}}\right)^{-2/3}.
\end{equation}
Nevertheless, for $\Psi < \Psi_{\rm in}$, where
\begin{equation}
\Psi_{\rm in} = \frac{4 \pi^2 v_{\rm in}^3\eta_0}{i_0\Omega_0^2},
\label{inin}
\end{equation}
the flow remains particle dominated: $E_{\rm n} \approx v_{\rm in}^2/2$.
The condition $\Psi_{\rm in} = \Psi_{0}$ just corresponds to
$\Omega_{\rm F} = \Omega_{\rm cr}$.
Remember that in the non-relativistic case the flow can
pass smoothly the critical surfaces only if $W_{\rm part}(r_{\rm
F}) \sim W_{\rm em}(r_{\rm F})$ \citep{Heyvaerts-96}. Thus, the flow at large
distances is to be particle dominated.

At first, let us consider the simplest case of sub Alfv\'enic cylindrical flow
${\cal M}^2 < 1$. Solving equations (\ref{11a}) and
(\ref{GS-Me}) one can find that the poloidal magnetic field
remains constant inside the jet up to the very boundary. Thus,
one can put $B(0) = B_{\rm ext}$. There is a simple physical
explanation why the homogeneous poloidal magnetic field is a
solution of the trans-field equation for a subsonic flow. The
point is that for ${\cal M}^2 < 1$ the energy density of the
poloidal magnetic field $B_{\rm p}^2/8\pi$ is much larger than
both the energy density of the toroidal magnetic field
$B_{\varphi}^2/8\pi$ and the energy density of particles
$\rho_{\rm m} v^2/2$. As a result, the trans-field equation can
be rewritten as
\begin{equation}
\frac{{\rm d}}{{\rm d}r}\,\left(\frac{B_{\rm p}^2}{8\pi}\right) = 0.
\end{equation}
Hence, for sub Alfv\'enic flows the homogeneous poloidal magnetic field
is a solution of the trans-field equation for arbitrary
invariants $E_{\rm n}(\Psi)$ and $L_{\rm n}(\Psi)$. But such sub
Alfv\'enic flow can exist only in the presence of large enough
external magnetic field $B_{\rm ext} > B(r_{\rm F})$, where
\begin{equation}
B(r_{\rm F}) = \left(\frac{\rho_{\rm in} v_{\rm in}^2}{B_{\rm in}^2/8\pi}\right)^{2/3}
\left(\frac{\Omega_0R_{\rm in}}{v_{\rm in}}\right)^{2/3}B_{\rm in}.
\label{i20}
\end{equation}
 For ordinary YSO $B(r_{\rm F}) \sim 10^{-1}$ G, so sub Alfv\'enic
jets cannot be realized.

On the other hand, cylindrical trans-Alfv\'enic flow can not also be realized
both for the center part of a flow or for the self-similar region. To proof
this proposition we must make two assumptions. We suppose that the
derivative of the Alfv\'enic Mach number remains finite at the
Alfv\'en surface (AS), i.e.
$\left.L-\Omega_{\mathrm{F}}r^{2}\right|_{\mathrm{AS}}=0$. We
also assume that the total current $I$ is not closing at the AS
strictly.

Let us suppose that the flow in the center of a cylindrical jet
is sub-Alfv\'enic and is about to cross the AS:
$M^{2}=1-\varepsilon$,
$L^{2}=\Omega_{\mathrm{F}}^{2}r^{4}-\delta$, where
$\varepsilon>0,\;\delta>0$. In this case one can write down the
leading terms of equation (\ref{GS-Me}) as
\begin{eqnarray}
\left(2e_{\mathrm{n}}+\Omega_{\mathrm{F}}^{2}r^{2}
\right)\frac{{\rm d}{\cal M}^{2}}{{\rm d}r}=\frac{1-\varepsilon}{\varepsilon}
\left[\Omega_{\mathrm{F}}^{2}r\varepsilon^{2}-\frac{\delta}{r^{3}}+\right.
\nonumber \\
\left.\varepsilon\frac{{\rm d}\Psi}{{\rm d}r}\left
(\frac{{\rm d}e_{\mathrm n}}{{\rm d}\Psi}+\frac{r^{2}}{2}\frac{{\rm d}\Omega_{\mathrm F
}^{2}}{{\rm d}\Psi}+\left(2e_{\mathrm n}
+\Omega_{\mathrm F}^{2}r^{2}\right)\frac{1}{\eta}\frac{{\rm d}\eta}{{\rm d}\Psi}\right)\right].
\label{1}
\end{eqnarray}
The term in r.h.s. part of equation is equal to zero for the
inner part of the flow, or it is negative for the self-similar
integrals. As the total current does not close at the AS,
\begin{equation}
\left.\frac{L-\Omega_{\mathrm{F}}r^{2}}{1-{\cal M}^{2}}=
\frac{\delta}{\varepsilon(L+\Omega_{\mathrm{F}}r^{2})}\right|_{AS}
\rightarrow\mathrm{const}\neq 0,
\end{equation}
i.e., $\delta=O(\varepsilon)$, we can neglect the first term
in r.h.s. bracket in (\ref{1}). This leads us to a conclusion
that the Mach derivative in the vicinity of the AS is negative.
However, if we assume that ${\cal M}^{2}$ should reach the unity, there
must be at least one point in the vicinity of the AS, where the
derivative is positive. We have come to a contradiction, so the
transition of the AS is impossible. One can easily prove by
analogy that the trans-Alfv\'enic flow is impossible also if the
flow is super-Alfv\'enic in the center.

Thus, super Alfv\'enic cold cylindrical flow must be supersonic at the
rotational axis: ${\cal M}_0^2 > 1$. In this case we return to the
jet-like solution \citep{eichler, bg1}
\begin{equation}
{\cal M}^{2} =
{\cal M}_{0}^{2}\left(1 + \frac{\Omega^2 r^2}{v_{\rm in}^2}\right).
\label{sol1''}
\end{equation}
But, as it was already stressed, in the presence of a finite
external pressure this solution is possible only if the central
core $r < r_{\rm core} = v_{\rm in}/\Omega$ contains almost all
magnetic flux $\Psi_0$. This can be realized only for a slow
rotation $\Omega_{\rm F} \ll \Omega_{\rm cr}$. In this case the
magnetic field on the axis cannot be much smaller than
\begin{equation}
B_{\rm min} = \frac{\Psi_0}{\pi r_{\rm core}^2}.
\end{equation}
Integrating now equations (\ref{11a}) and (\ref{GS-Me}), one can obtain that
\begin{equation}
B_0 \approx \frac{B_{\rm min}}{\ln(1 + B_{\rm min}/B_{\rm ext})}.
\label{B0min}
\end{equation}
Accordingly,
\begin{equation}
\Psi_{\rm core} \approx \frac{\Psi_0}{\ln(1 + B_{\rm min}/B_{\rm ext})}.
\end{equation}
This structure was reproduced numerically as well \citep{Lery-99}.

But for fast rotation $\Omega_{\rm F} \gg \Omega_{\rm cr}$ the
reasonable solution cannot be realized as the core magnetic flux
$\Psi_{\rm core}$ is much smaller even than the flux $\Psi_{\rm
in}$ (\ref{inin}) within the central part of a flow. Indeed, as
according to definitions (\ref{4a}) and (\ref{10a}) one can write
down
\begin{equation}
B_{\rm p}(0) = \frac{4 \pi \eta_{\rm n}}{{\cal M}_0^2},
\end{equation}
we obtain for ${\cal M}_0^2 > 1$
\begin{equation}
\frac{\Psi_{\rm core}}{\Psi_{\rm in}}
\approx \frac{i_0 v_{\rm in}}{2c  M_{0}^{2}}
< \left(\frac{\Omega_{\rm F}}{\Omega_{\rm cr}}\right)^{-2/3} \ll 1.
\end{equation}
It means that the cold cylindrical flow resulting from the
interaction of supersonic, fast rotating wind with external media
cannot be realized.

\subsubsection{Heating at the oblique shock}

To resolve this contradiction, one can propose that in the
observed non-relativistic supersonic jets an important role in the force
balance may play the finite temperature. E.g., additional heating
can be connected with the oblique shock near the base of a jet
\citep{bt, Levinson-07}. It is well known that such a shock is
needed to explain the emission lines observed in jets
from YSO \citep{Schwartz-83}. This situation is alike the pure
hydrodynamical supersonic outflow meeting the wall. The
hydrodynamical analogy is all the more reasonable as the
non-relativistic supersonic outflow is to be particle dominated.

To evaluate the thermal terms in equations (\ref{11a}) and
(\ref{GS-Me}) we consider pure hydrodynamical shock wave turning
the spherically symmetric supersonic flow into the cylindrical
jet (see \citep{Grenoble} for more detail).
For the field lines of the pre-shock flow at the inclination
angles to the rotational axis less than the critical one, we
seek the shock position so as to turn the flow into a cylinder.
The critical angle for the pre-shock sound Mach number
$M_{\mathrm s, 1}^{2}\gg 1$ is equal to
\begin{equation}
\theta_{\rm max} = \sin^{-1}(1/\gamma),
\end{equation}
where $\gamma$ is the polytropic index. In particular, for
$\gamma=1.2$ we have $\theta_{\rm max}\approx 56^{\circ}$. For the rest
field lines we model the shock position to transit smoothly from
the $\theta=\theta_{\rm max}$ to the equatorial field line $\theta=\pi/2$.
Knowing now the shock position, we can calculate the entropy jump
$\Delta s$ for every field line. The rest integrals of motion, according
to conservation laws, are to be unbroken on the oblique shock.

Thus, we can solve one-dimensional equations (\ref{11a}) and (\ref{GS-Me})
taking into account the effects of the heating on a shock through the
corresponding thermal terms. We find that for the super
Alfv\'enic flow the scaling $\Psi\propto \ln \, r$ holds no more, so
the jet boundary is located at the finite distance from the
rotational axis. Obtained jet parameters for typical TTauri star
(a jet radius $R_{\rm jet}\sim 10^{15}$ cm, a temperature behind
a shock needed to give rise to the forbidden emission lines
$T\sim 10^{4}$ K, a poloidal velocity $v_{\rm p}\sim 10^{7}\div
3\cdot 10^{7}$ cm/s, and a toroidal velocity $v_{\rm \varphi}\sim
10^{6}$ cm/s) are in agreement with the observational data.

\subsubsection{In the center of the self-similar solution}

For non-relativistic cold outflow the self-similar invariants are
\citep{bp}
\begin{eqnarray}
\Omega_{\rm F}(\Psi) & = &  \Omega_0 (\Psi/\Psi_{\rm b})^{-3\beta/2},
\label{int1bis} \\
E_{\rm n}(\Psi) & = & E_0 (\Psi/\Psi_{\rm b})^{-\beta},
\label{int2bis} \\
L_{\rm n}(\Psi) & = & L_0 (\Psi/\Psi_{\rm b})^{\beta/2},
\label{int3bis} \\
\eta_{\rm n}(\Psi) & = & \eta_0 (\Psi/\Psi_{\rm b})^{1 - 3\beta/2}.
\label{int4bis}
\end{eqnarray}
Again, one can seek the solution of the two-dimensional GS equation for
$\Psi > \Psi_{\rm b}$
in the form $\Psi(R,\theta) = R^{1/\beta} \Theta(\theta)$, so that for
$\theta \ll 1$
\begin{equation}
\Psi(R,\theta) = {\cal A} R^{1/\beta} \theta^a.
\label{autoan'}
\end{equation}
Then, the cylindrical radius of the boundary $\Psi = \Psi_{\rm b}$
can be written as
\begin{equation}
r_{\rm b}(z) = {\cal A}^{-1/a} \Psi_{\rm b}^{1/a} z^{1 - 1/a\beta}.
\end{equation}

As a result, integrating one-dimensional cylindrical equations (\ref{11a}) and
(\ref{GS-Me}) in the region $\Psi < \Psi_{\rm b}$ for particle
dominated flow, i.e., using the solutions (\ref{sol1'})--(\ref{sol1}), we obtain
for $\Omega_{\rm F}  =\Omega_0$ and $\eta =\eta_0$
\begin{equation}
{\cal M}^2_{\rm b}(z)
= \frac{8\pi^2 \eta_0 \mu}{a R_{\rm L}{\cal A}^{3/a} \Psi_{\rm b}^{1-3/a}} z^{3 -3/a\beta}.
\end{equation}
On the other hand, for $\Psi > \Psi_{\rm b}$ ($r > r_{\rm b}$)
one can seek again the solution in a form
${\cal M}^2(r) = {\cal M_{\rm b}}^2(r/r_{\rm b})^{b}$.
As a result, equations (\ref{11a}), (\ref{GS-Me}) give
\begin{equation}
a = 2, \quad b = 2 - 4\beta.
\label{asy3}
\end{equation}
Substituting now $r = R \theta$, we see that
\begin{equation}
{\cal M}^2 = {\cal C} \theta^{\varepsilon}.
\end{equation}
Again, the coefficient
${\cal C} \propto {\cal M_{\rm b}}^2(R) R^{b}/r_{\rm b}^{b}(R)$,
in agreement with self-similar property, does not depend on $R$.

The results of numerical integration of the system (\ref{11a}),
(\ref{GS-Me}) for the cold non-relativistic particle dominated
flow are presented in Figs. 3. Here we also see very good
agreement between numerical results and analytical asymptotic
behaviour (\ref{asy3}). The rapid growth of the exponent $b$
close to the end of the calculation is due to drop of Mach number
close to unity where the flow is close to the Alfv\'enic surface.
However, as we have showed, the smooth transition of the AS is
impossible for the chosen self-similar integrals, so we must stop
our calculation at this point. The power $b = 2 - 4\beta$ was
confirmed by \citet{vla} as well.

\begin{figure}
\includegraphics{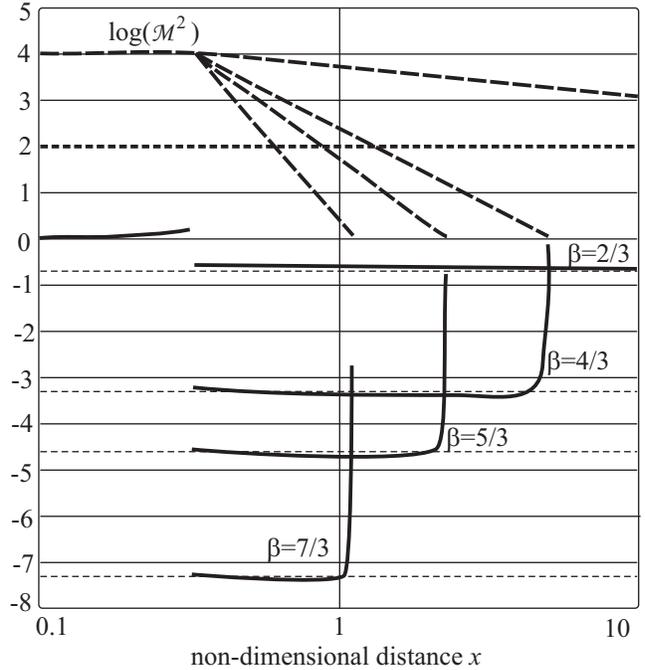}
\caption{Internal structure of particle dominated
non-relativistic jet. The short dashes line is an exponent $a$ of
the flux function $\Psi$. The long dashes line represents the
$\mathrm{log}{\cal M}^2$, and the solid line represents an
exponent $b$ of ${\cal M}^2$ for different values of the
parameter $\beta$. The thin dashes lines are the theoretical
exponents $b = 3 - 6\beta$. The non-dimensional radius
$x=\Omega_{\rm F}(0)r/v_{\rm in}$.}
\end{figure}

\section{Conclusion}

Thus, the cylindrical Grad-Shafranov equation has definite
advantages in comparison with the standard self-similar ones.
Using this approach it was demonstrated that in the relativistic
case an effective particle acceleration can take place only if
$r_{\rm jet} \geq \sigma R_{\rm L}$, the curvature of magnetic
surfaces playing no role. We found as well that for
non-relativistic flows which are magnetically dominated near the
origin the solution can be constructed only in the presence the
oblique shock near the base of a jet where the additional heating
is to take place. In all cases the magnetic flux within the
central core was determined. As was demonstrated, for
relativistic flow the central core is to contain only a small
part of the total magnetic flux. For the non-relativistic outflow
the situation is opposite.

\section{Acknowledgments}

We thank Prof. A.V.Gurevich for his interest and support,
G.Pelletier, J.Ferreira, and N.Vlahakis for useful discussions.
This work was partially supported by Russian Foundation for Basic
Research (Grant no.~08-02-00749), the Landau Scholarship provided
by the Juelich Institute for Solid State Physics, and the Dinasty
fund.

{}

\end{document}